\documentclass[sigconf,nonacm]{acmart}

\AtBeginDocument{%
  \providecommand\BibTeX{{%
    \normalfont B\kern-0.5em{\scshape i\kern-0.25em b}\kern-0.8em\TeX}}}

\begin{document}

\title{FractalBrain: A Neuro-interactive Virtual Reality Experience using Electroencephalogram (EEG) for Mindfulness}
\renewcommand{\shorttitle}{FractalBrain: Neuro-interactive VR using EEG for Mindfulness}

\author{Jamie Ngoc Dinh}
\email{ngocdinh@umd.edu}
\orcid{0009-0007-1100-556X}
\affiliation{%
  \institution{University of Maryland,\linebreak College Park, USA}
  \city{}
  \state{}
  \country{}
}

\author{You-Jin Kim}
\email{yujnkm@ucsb.edu}
\orcid{0000-0003-0903-8999}
\affiliation{%
  \institution{University of California,\linebreak Santa Barbara, USA}
  \city{}
  \state{}
  \country{}
}
\author{Myungin Lee}
\email{myungin@umd.edu}
\orcid{0000-0002-4202-4364}
\affiliation{%
  \institution{University of Maryland,\linebreak College Park, USA}
  \city{}
  \state{}
  \country{}
}

\renewcommand{\shortauthors}{Dinh et al.}

\begin{abstract}
Mindfulness has been studied and practiced in enhancing psychological well-being while reducing neuroticism and psychopathological indicators. However, practicing mindfulness with continuous attention is challenging, especially for beginners. In the proposed system, FractalBrain, we utilize an interactive audiovisual fractal with a geometric repetitive pattern that has been demonstrated to induce meditative effects. FractalBrain presents an experience combining a surreal virtual reality (VR) program with an electroencephalogram (EEG) interface. While viewing an ever-changing fractal-inspired artwork in an immersive environment, the user's EEG stream is analyzed and mapped into VR. These EEG data adaptively manipulates the audiovisual parameters in real-time, generating a distinct experience for each user. The pilot feedback suggests the potential of the FractalBrain to facilitate mindfulness and enhance attention.

\smallskip
\textit{ This is a preprint version of this article. The final version of this paper can be found in the Extended Abstracts (Interacticity) of ACM CHI 2024. For citation, please refer to the published version.}
\textit{This work was initially made available on the author's personal website [yujnkm.com] in April 2024, and was subsequently uploaded to arXiv for broader accessibility.}

\end{abstract}

\ccsdesc[500]{Human-centered computing~Virtual reality}
\ccsdesc[500]{Applied computing~Media arts}
\ccsdesc[300]{Applied computing~Sound and music computing}

\keywords{Virtual Reality, Audiovisual, EEG, Fractal, Interactive Technologies, Audiovisual, Neurofeedback, Mindfulness}

\begin{teaserfigure}
  \centering
  \includegraphics[width=0.9\textwidth]{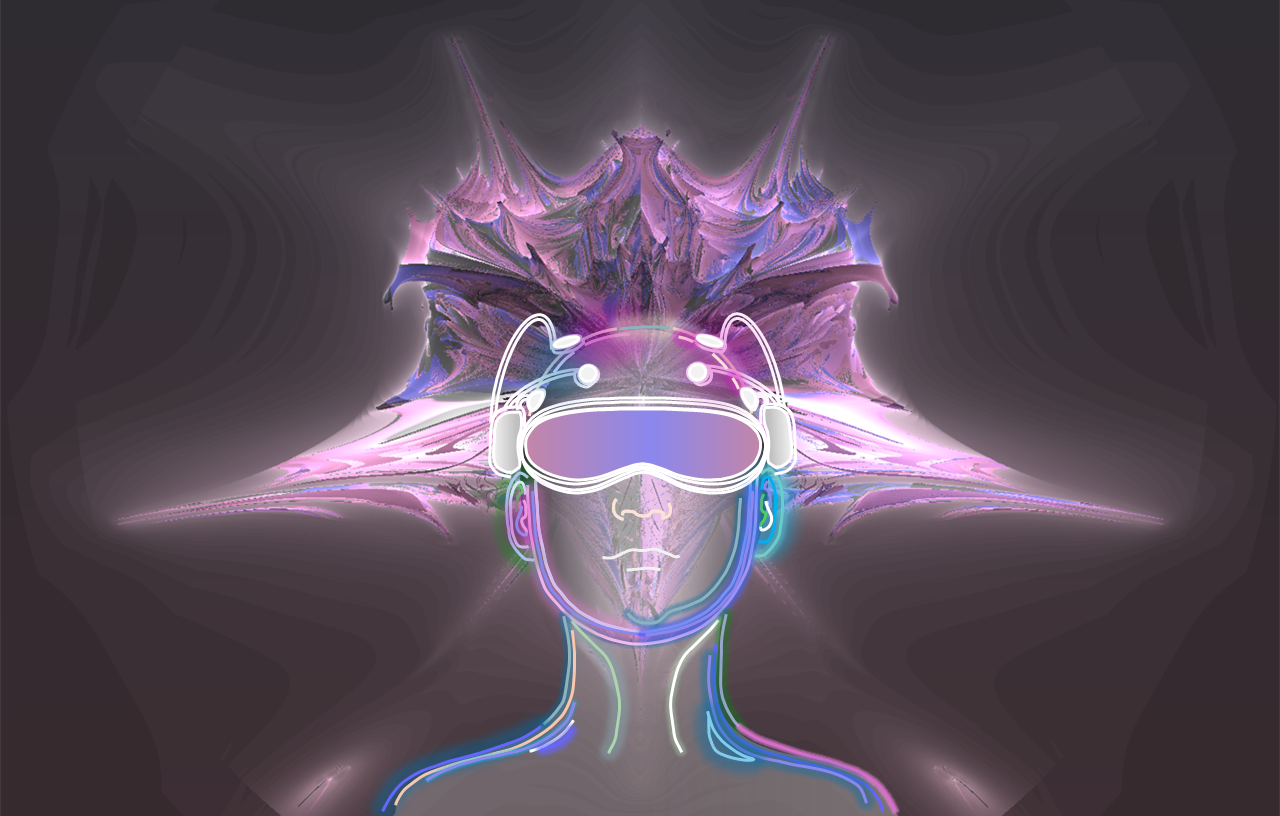}
  \caption{Conceptual image of FractalBrain: A Neuro-interactive VR Experience using EEG for Mindfulness.}
  \Description{Conceptual image of FractalBrain with the fractal-inspired artwork and an illustration of a user wearing a VR and EEG headset.}
  \label{fig:teaser}
\end{teaserfigure}


\maketitle

\section{Introduction}
Mindfulness is a type of meditation that brings one's complete attention to the present experience on a moment-to-moment basis without interpretation or judgment \cite{marlatt99}. Research has shown that mindfulness induces a range of positive psychological outcomes, such as improved subjective well-being, decreased psychological symptoms and emotional reactivity, and enhanced behavioral regulation~\cite{keng}. In recent years, the rapid development of electroencephalogram (EEG) technology has allowed new observations and interpretations of mindfulness. However, maintaining continuous attention is a challenging skill to achieve for the novices~\cite{lutz}, and experiencing mindfulness is becoming rare. 

Given this background, we propose a novel experience to answer the question: "Can we achieve a mindfulness experience using digital mediation with evolving technology?" Our project, FractalBrain, aims to promote mindfulness by utilizing an audiovisual fractal to create a neuro-interactive experience that combines a surreal virtual reality (VR) experience with an EEG interface. The EEG interface analyzes the user's mental status and controls the dynamic audiovisual fractal in an immersive environment in real-time.  

\subsection{VR and EEG}
The emerging EEG and VR technology with high computation and resolution has allowed new methods to derive meditative effects. For example, commercial VR games that offer various virtual environments for meditation have increased popularity and attention to the practice~\cite{Meditation2021VR} \cite{maloka}. 
Lin (2023) presented a VR artwork visualizing EEG and heart rate data relating the human body to cosmic nebulae \cite{lin2023body}. Lin (2021) employed EEG signals to generate phyllotaxis in the form of installation art, demonstrating its meditative effects through a study involving 50 participants \cite{lin}. Kosunen (2016) investigated the combination of VR and EEG, creating a neuroadaptive meditation system where real-time EEG theta and alpha band power are translated into levitation and visual effects in the VR environment \cite{kosunen}. While these approaches have effectively supported mindfulness meditation, the majority of them employed natural settings and static scenery \cite{chandrasiri} \cite{flores}. Moreover, EEG is used in mixed reality to measure cognitively relevant signals \cite{krugliak} and to facilitate natural locomotion \cite{investigating, kumaran2023impact}.

\subsection{Fractal}
Our project presents ever-changing, fractal evolution-inspired interactive artwork in VR. Fractal, introduced by Benoit Mandelbrot in 1975, is characterized by a repetitive pattern that recurs on progressively smaller scales, forming shapes of significant intricacy \cite{mandelbrot}. Several researches have pointed out that fractals have the potential to induce a sense of relaxation and meditative effects. In an experiment performed at NASA-Ames Research Center, 24 participants exposed to diverse fractal images exhibited lower stress levels during mental tasks, as indicated by skin conductance monitoring \cite{taylor06}. Extended on this study, using recorded continuous EEG in 32 participants, Taylor (2011) showed that these fractals elicited a maximal alpha response in the frontal region, indicating the induction of a relaxed state \cite{taylor11}. With a similar study utilizing EEG, Hagerhall (2008) showed that fractals with balanced visual complexity likely activate most in the parietal region, suggesting potential efficiency in capturing attention \cite{hagerhall}. 

While conventional studies suggest the potential of each component, we present a new design approach to create an immersive, interactive, and personalized experience using VR, EEG, and fractal to aid in practicing mindfulness.

\begin{figure}
\includegraphics[width=0.58\columnwidth]{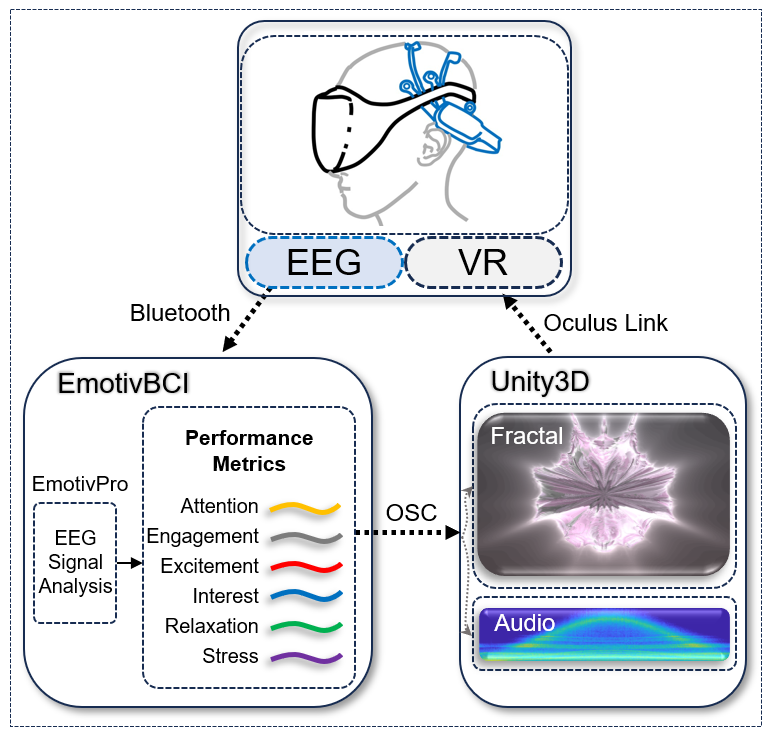}
\includegraphics[width=0.39\columnwidth]{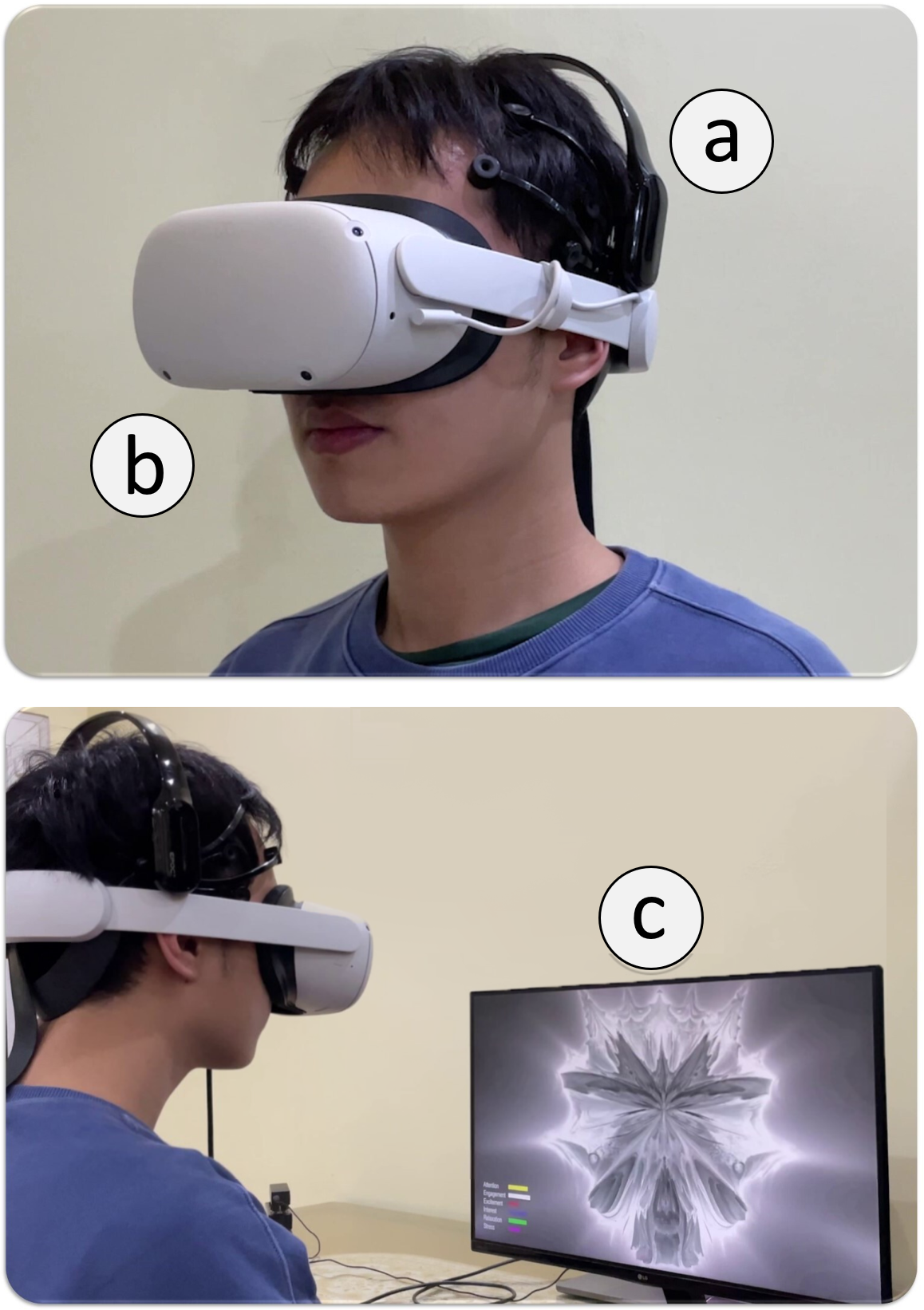}
 \caption{Schematic of the FractalBrain (left) and the demonstration (right). (a) Emotiv EEG headset. (b) Meta Quest 2 VR headset. (c) Display mirroring user's view from the VR headset.}
 \label{fig:system}
 \Description{Schematic of the FractalBrain system (left) and the real demonstration system (right). The system includes an Emotiv EEG Headset, a Meta Quest 2 VR headset, and a display mirroring user's view from the VR headset.}
\end{figure}

\section{The neuro-interactive System Design}
The system involves VR for audiovisuals and EEG for the interaction. For VR, Unity3D \cite{unity} game engine and programming language C\# is used to develop the application. In this work, we use Meta Quest 2 as the VR headset, but our project is compatible with various head-mounted displays and not restricted to specific devices. For EEG, we use the Emotiv EPOC X \cite{emotiv}, a 14-channel wireless EEG headset. Emotiv EPOC X has demonstrated reliability for non-clinical applications \cite{duvinage}, and has been utilized in various research projects \cite{yueliu} \cite{muhammad} \cite{malete} \cite{kumar}. The EEG headset and software packages, including EmotivBCI and EmotivPro, can capture and analyze users' brainwaves and stream the raw EEG data, band power, and performance metrics ($Attention, Engagement, Excitement, Interest, Relaxation$, and $Stress$). 

Figure ~\ref{fig:system} shows the schematic of our system. At first, the EEG headset measures the brain signal from 14 channels and transmits its data to the EmotivBCI software via Bluetooth. The EmotivPro software analyzes the signal and extracts the performance metrics. These metrics are then sent to Unity3D through Open Sound Control (OSC) \cite{osc}, manipulating the fractal and audio parameters in Unity3D. To avoid abrupt changes in the scenery, an adaptive algorithm gradually shifts the parameter. The VR headset with Oculus Link wirelessly delivers the stereo audio and 3D visuals from Unity3D. Repetitively, when the experience evokes a change in the EEG metrics, the parameters of fractal and audio change to compensate for the user's status. 

Within the same generative fractal algorithm, by observing the user's status and applying it back, the experience immerses the users with counterpoints by balancing predictability and uncertainty. For example, when the user's attention gets low, the fractal morphs more compellingly to arouse additional curiosity; if the user gets too excited or stressed, the system reduces the fractal's dynamic. Such cognition can be different for each user because the degree of perception differs for individuals. The proposed interface personalizes the fractal's morphology with EEG data to articulate an engaging interactive experience, preventing it from being monotonous. Furthermore, based on the observation that our brain synthesizes information from cross-modal stimuli through multisensory integration \cite{Stein08}, we designed our VR experience to be audiovisually coherent to obtain natural stimuli through digital mediation. This design makes the experience feel more natural, even with the surreal digital fractal.

\begin{figure}[t]
\centering \includegraphics[width=\columnwidth]{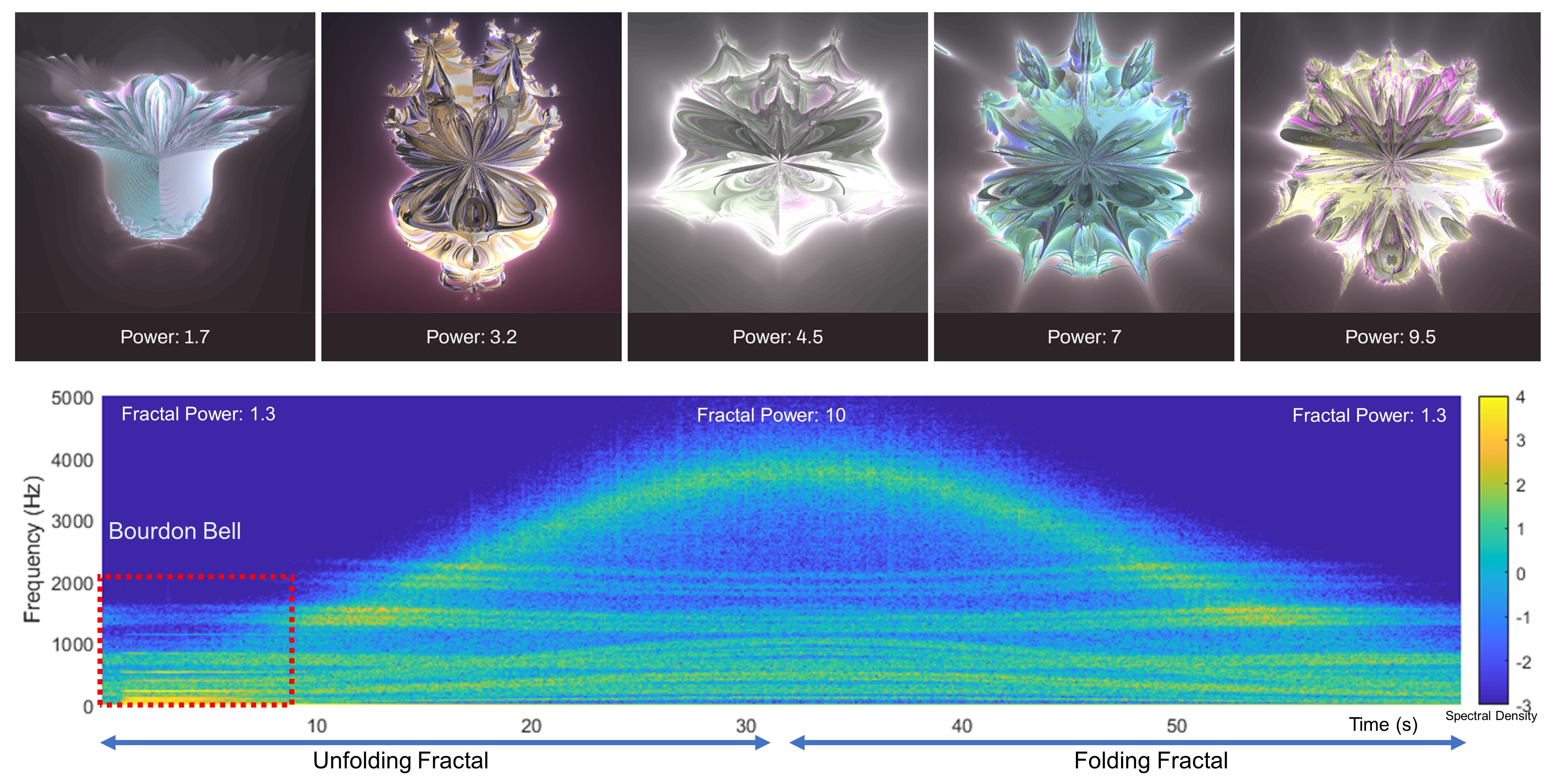}
 \caption{Exemplary visual instances with various fractal parameters (up) and audio spectrogram of FractalBrain's phrase: unfolding and folding structure over time (down).}
 \label{fig:variances}
 \Description{Up: Exemplary visual instances with various fractal parameters including colors and power ranging from 2 to 10. Down: Audio spectrogram of FractalBrain's phrase: unfolding and folding structure over time relative to fractal power.}
\end{figure}

\subsection{Visual Rendering} %
To present an artistic visual experience in VR, we developed a fractal-inspired graphic in a third dimension using a custom ray marching engine. Ray marching is a rendering technique to produce real-time graphics using modern GPUs \cite{hart96}. The method enables complex 3D geometry, such as fractals, to be rendered efficiently \cite{perlin89, hart89}. 

Our visual utilizes variations of Mandelbulb, which is the extension of the Mandelbrot fractal set in 3D, with a basic formulation as described below \cite{daniel}:

For a 3D point's polar coordinate of x, y, z:

\begin{center}\((x,y,z)^n = r^n [sin(\theta*n) * cos(\phi*n) , sin(\theta*n) * sin(\phi*n) , cos(\theta*n)]\)\end{center}

where $r = \sqrt{x^2 + y^2 + z^2}$, $\theta = atan2(\sqrt{x^2 + y^2}, z)$, and $\phi = atan2(y,x)$.

Utilizing the above basic formulation, we have twisted the variable \(\theta\) to \(\theta = acos(z*x/r)\). The equation for the 3D point's polar coordinate is modified as follow:

\begin{center}\((x,y,z)^n = r^n [(1 - bW) * sin(\theta) * cos(\phi) + bW * sin(\theta) - cos(\phi), sin(\phi) * sin(\theta), cos(\theta)]\)\end{center}

where $bW$ is the color index.

The formula is then inserted into a ray marching engine to render the graphic. Multiple parameters control the fractal, and the system maps the EEG metrics with the following relation: 
\begin{itemize}
\item Power: the visual complexity of the fractal, with higher values indicating intricate details and delicate structure, while lower values result in smoother and sparser shapes. \newline
$Power \propto Attention + Excitation^{-1}$
\item Color: the color mix of the fractal \newline
$Red \propto Excitation^{-1} + Engagement^{-1},\quad Green \propto Attention^{-1} + Stress^{-1}, \quad Blue \propto Relaxation^{-1} +  Interest^{-1}$
\item Power Increase Rate: ascending and descending quantity of the fractal power per frame. $\Delta Power \propto Excitation^{-1}$
\item Oscillation Rate: the speed at which the fractal runs through one complete loop. $f_{loop} \propto Excitation^{-1} + Stress^{-1}$
\end{itemize}

Figure ~\ref{fig:variances} (up) shows the exemplary instances of the graphic when changing the parameters.

\subsection{Audio Synthesis}
The audio is a fully interactive component of the project. We wrote a custom code in C\# to synthesize coherent audio using fractal and EEG parameters in real-time. To depict the morphing and reiterative nature of the fractal in the auditory domain, we designed an interactive stereo synthesizer motivated by Bourdon bells, ocean waves, and György Ligeti's Lux Aeterna, used in Stanley Kubrick's 2001: A Space Odyssey \cite{patterson2004music}. Figure ~\ref{fig:variances} (down) shows the exemplary audio spectrogram of the FractalBrain's phrase. Triggered by the Bourdon bell-like sound, 16 different frequency-modulated oscillators sculpt the unique timbre following the fractal parameters controlled by EEG metrics. This structure unfolds and folds over the phrase, which can differ from 50 to 70 seconds according to the EEG metrics. As if we hear and see the ocean wave fold and unfold, the audiovisual fractal presents a coherent digital experience with the proposed interface.

\section{Demonstration Procedure}
For our demonstration, a 14-channel wireless EEG headset will be used to monitor participants' brain activity. To ensure a safe and hygienic experience, multiple sets of pads will be available, cycled, and thoroughly cleaned before reuse. Each EEG sensor pad will have a small amount of a salt-based mineral solution liquid applied, and participants may feel a slight wet sensation on their scalps at times. Verbal instructions and consent will be obtained before placing the EEG headset on, ensuring participants are informed and comfortable.

Prior to each session, attendees will undergo a brief and non-intrusive setup process, guided by our team to ensure a seamless experience. During the demonstration, we will emphasize that EEG data is not recorded for privacy purposes, and the live view will be streamed for demonstration purposes only. The visual representation of the EEG data and live VR view will be shared on a large LCD screen next to the table. Verbal consent will be obtained from participants regarding the use and sharing of abstract visualizations of their EEG data during the session. Participants will have the option to remove the VR headset and EEG sensors at any time if they feel uncomfortable, prioritizing their comfort and consent throughout the experience.

\section{Pilot study and future work}
We conducted a pilot study with several users to gather feedback on the overall experience. Subjective responses revealed that participants felt they had been transported to another space distinct from the real world. Users reported increased attention to the experience in VR, with certain participants describing a sense of entering a meditative state of emptiness. Participants highlighted actively reflecting on their mental state to anticipate scenery changes. Almost everyone found this experience to be both enjoyable and impressive. These subjective responses align with the EEG observation, as nearly all participants exhibited significantly higher levels of $Attention$, $Engagement$, and $Interest$ compared to $Excitement$, $Relaxation$, and $Stress$. 

For future work, we aim to conduct a quantitative user study to examine the impact of the experience on users' mental states with higher-resolution EEG devices. With the brain interface, the goal is ultimately to understand the user's status and adaptively apply the corresponding response to achieve mindfulness and meditative effects. 

\section{Conclusion}
In this work, we presented FractalBrain: a neuro-interactive experience that combines an audiovisually surreal VR experience and EEG interface to foster mindfulness. Users are invited to a dynamic fractal-inspired audiovisual in VR while an EEG headset interprets the user's brain. The analyzed metrics manipulate the audiovisual parameters in real-time. The interaction mapping is designed based on narrative compositional theory to achieve mindfulness by compensating the user's status over time. 
The interactive EEG with various time-varying parameters allows a unique experience for every participant in each session, creating fresh and captivating encounters. Pilot feedback has shown positive outcomes regarding the potential of FractalBrain to enhance concentration and facilitate mindfulness practices.

\begin{acks}
This work is supported by the Immersive Media Design Program and the Department of Computer Science at the University of Maryland, College Park.
\end{acks}

\bibliographystyle{ACM-Reference-Format}
\bibliography{FractalBrain}


\end{document}